\documentclass{aa}
\usepackage{graphicx}        
\usepackage{psfig}        

\begin{document}

%
   \title{Density waves in the shearing sheet }
   \subtitle{V. Feedback cycle for swing amplification \\ by 
    non--linear effects}

   \author{B. Fuchs, C. Dettbarn, and T. Tsuchiya\thanks{{\emph Present 
   address:} SGI Japan Ltd., Ebisu Garden Place Tower 31F,
   4-20-3 Ebisu, Shibuya-ku, Tokyo 150-6031, Japan}}


   \institute{Astronomisches Rechen--Institut am Zentrum f\"ur Astronomie der
   Universit\"at Heidelberg, M\"onchhofstrasse 12--14, 69120 Heidelberg,
   Germany}

   \date{Received 2005; accepted 2005}

   \abstract{
   Non--linear effects in the dynamical evolution of a shearing sheet made of
   stars are studied. First the implications of hitherto neglected 
   non--linearities of the Boltzmann equation for the dynamical evolution of 
   the shearing sheet are investigated. Using a formalism developed 
   previously on the basis of the linearized Boltzmann equation it is 
   demonstrated that the inclusion of the non--linear term leads to a
   feedback cycle for swing amplified density waves in the  unbounded 
   shearing sheet. Such a feedback is unique to star disks and is not known for 
   gas disks. In order to present concrete examples of the non--linear feedback
   cycle a SCF code was developed and numerical simulations of the dynamical 
   evolution of the shearing sheet are performed. The numerical results
   seem to confirm the theoretical predictions. The evolution of the 
   shearing sheet resembles closely and might actually explain the recurrent 
   spiral instabilities found in large--scale numerical simulations of the 
   dynamical evolution of galactic disks.
      }
		
   \mail{fuchs@ari.uni-heidelberg.de}
    		
   \maketitle

%

\section{Introduction}

The shearing sheet (Goldreich \& Lynden--Bell 1965, Julian \& Toomre 
1966) model has been developed as a tool to study the dynamics of
galactic disks and is particularly well suited to theoretically describe
the dynamical mechanisms responsible for the formation of spiral arms.
For simplicity, the model describes only the dynamics of a 
patch of a galactic disk. It is assumed to be infinitesimally thin and 
its radial size is assumed to be much smaller than the disk. Polar 
coordinates can be therefore converted to pseudo-Cartesian
coordinates and the velocity field of the differential rotation of the
disk can be approximated by a linear shear flow. These simplifications
allow an analytical treatment of the problem, which helps to clarify 
the underlying physical processes operating in the disk. In the previous
papers of this series (Fuchs 2001a, b, 2004, 2005) we have studied various 
aspects of the dynamics of spiral density waves in a shearing sheet of 
stars based on the {\it linearized} Boltzmann equation. 
In the present paper we study {\it non--linear} effects in the dynamical
evolution of the shearing sheet. In particular, we investigate the role of the
hitherto neglected non--linear term of the Boltzmann equation. The principal 
result is that the non--linearity leads to a feedback cycle for swing
amplification of spiral density waves. Recurrent spiral instabilities, 
which lead to an ever changing appearance of the disk, have been 
reported by Sellwood \& Carlberg (1984) and Sellwood (1989) from their
numerical simulations of the dynamical evolution of self gravitating, 
differentially rotating disks. Toomre (1990) and Toomre \& Kalnajs (1991) have 
argued that these recurrent spiral density waves are essentially due to swing
amplified (Toomre 1981) random fluctuations of the surface density of the disks.
Sellwood (1989), on the other hand, points out that contrary to expectation the
spiral instabilities stay at the same level of activity in the experiments, 
if the particle number is increased and thus the particle noise is reduced. 
Moreover, Sellwood \& Carlberg (1984) have shown that, once the cycles 
of recurrent transient spiral density waves have established themselves in 
their simulations, the leading density wavelets which are then swing amplified 
are extremely unlikely due to chance alignments 
of the stars, because the amplitudes of these leading wavelets are too large.
They consider the recurrent spirals not to be unconnected random events 
arising from swing amplified random noise. Sellwood (2000) puts
forward the notion of a feedback mechanism for spiral instabilities, which is 
related to fine structure of the distribution function of stars in phase space
carved in by transient spiral density waves, which can then in turn incite 
density waves again. As a concrete example Sellwood \& Lin (1989) and Sellwood 
\& Kahn (1991) have developed the concept of recurring groove instabilities. 
These rely on fairly sharp boundaries of the dynamically active parts of the 
self gravitating disks, because they are initiated by edge modes (Toomre 1981). 
The groove instabilities develop then as rigidly rotating spiral modes of the  
disk\footnote{ It can be shown (Toomre \& Kalnajs 1980, unpublished results,
Sellwood \& Kahn 1991, Fuchs 2005) that even the shearing sheet develops rigidly
propagating, exponentially growing spiral modes, if a sharp left hand boundary
is introduced.}and
seed fresh grooves through particle wave resonance at their Lindblad 
resonances. However, Sellwood \& Carlberg (1984) have shown in detail 
that the spiral structures seen in their numerical simulations are transient 
swing amplified shearing density waves exactly of the kind found in the 
infinite shearing sheet model. They demonstrate not only that the spiral arms 
shear from leading to trailing orientation, but also that the density 
waves grow preferentially with azimuthal wave numbers predicted by the shearing 
sheet model, which has become widely known
as the Toomre (1981) ``$X=2$'' prescription. Exactly the same was found by
Fuchs \& von Linden (1998) in their numerical simulations of the dynamical
evolution of a self gravitating disk made of stars and interstellar gas. Toomre
(1990) has argued convincingly that the spiral arms of Sc galaxies are such 
transient shearing density waves. One
of the deeper reasons for this success of the infinite shearing sheet model in 
describing spiral density waves realistically is the rapid convergence of the 
Poisson integral in self gravitating disks (Julian \& Toomre 1966). Consider, 
for example, the potential of a sinusoidal density perturbation
\begin{equation} 
\Phi(x,y) = - G \int_{-\infty}^{\infty} dx' \int_{-\infty}^{\infty} dy'
\frac{\Sigma_{10} \sin{(kx')}}{\sqrt{ (x-x')^2 + (y-y')^2}} \,,
\end{equation} 
where $G$ denotes the constant of gravitation. At $y=0$
\begin{eqnarray} 
&& \Phi = - 4G \Sigma_{10} \sin{(kx)} \lim_{ x_{\rm L} \to 
\infty} \frac{{\rm Si}(k x_{\rm L})}{k} = \nonumber \\ &&
 - \frac{2 \pi G \Sigma_{10}\sin{(kx)}}{k}\,.
\end{eqnarray}
The sine integral in equation (2) converges so rapidly that at $k x_{\rm L} 
= \frac{\pi}{2}$ it reaches already 87\% of its asymptotic value. Thus
the ``effective range'' of gravity is only about a quarter of a wave length. 
The shearing sheet models effectively patches of galactic disks of such size.
Thus the shearing sheet model seems to be a well suited tool to interpret the
recurring transient spiral instabilities seen in the numerical simulations of
self gravitating disks. Of course, the shearing sheet model with its rather
drastic simplifications such as the neglected curvature of the mean guiding 
centers of the stellar orbits and the assumed homogeneous mass distribution 
or the periodic boundary conditions in numerical simulations of the dynamical
evolution has its limitations. Sellwood \& Carlberg
(1984) noted in particular that the maximum growth factors of the shearing
density waves in their simulations are somewhat larger then the predictions of
the local model. Similarly Toomre (1981) finds also greater amplification of
density waves in the full disk model of Zang.

It is the aim of 
the present paper to demonstrate on the basis of the infinite shearing sheet 
model that a feedback cycle of swing amplified density waves is indeed 
theoretically expected. In the next section a formal derivation is presented, 
while in the third section we describe numerical SCF simulations 
of the dynamical evolution of the shearing sheet and illustrate the feedback
cycle with concrete examples. The SCF simulations are directly compared 
with N--body simulations of the dynamical evolution of the shearing sheet by 
Toomre (1990), Toomre \& Kalnajs (1991) and Chiueh \& Tseng (2000).

Non--linear interaction of density waves has been previously
studied by Tagger et al.~(1987) and Sygnet et al.~(1988). These authors 
considered, however, the interaction of rigidly rotating, slowly growing spiral
modes, which do not develop in the unbounded shearing sheet. A preliminary 
report on the work presented here was given in Fuchs (1991).

\section{Theoretical considerations}

\subsection{Boltzmann equation}

The time evolution of the distribution function of stars in phase space, $f$,
is determined by the collisionless Boltzmann equation
\begin{equation}
\frac{\partial f}{\partial t} + \left[f, H\right] = 0\,,
\end{equation}
where the square bracket indicates the usual Poisson bracket and $H$ is
the Hamiltonian of the stellar orbits (cf.~paper I).
Spiral arms are usually thought to be only minor perturbations of 
galactic disks. Thus we choose as in the previous studies a
perturbation `Ansatz' of the form
\begin{equation}
f = f_{\rm{0}} + \delta f \,, \; H = H_{\rm{0}} + \delta \Phi\,,
\end{equation}
where $f_{\rm 0}$ and $H_{\rm{0}}$ refer to the unperturbed shearing 
sheet, respectively. The physical idea is to subject the shearing 
sheet to a small spiral--like potential perturbation $\delta \Phi$ and 
to determine the disk response $\delta f$ by solving the Boltzmann 
equation. Inserting the `Ansatz' (4) into the Boltzmann equation (3) 
leads to 
\begin{equation}
\frac{\partial \delta f}{\partial t} + \left[f_{\rm{0}} , \delta \Phi
\right]+ \left[ \delta f , H_{\rm{0}} \right] = - [\delta f, \delta 
\Phi]\,,
\end{equation}
where the rhs represents the non--linear term of the Boltzmann equation.
The background distribution function $f_{\rm 0}$ is chosen in the following,
again in the form of a Schwarzschild distribution function.

The non--linear Poisson bracket has the explicit form
\begin{eqnarray}
&& [\delta f, \delta \Phi] = \nonumber \\ &&
\frac{\partial \delta f}{\partial w_1}
\frac{\partial \delta \Phi}{\partial J_1} + \frac{\partial \delta f}
{\partial w_2} \frac{\partial \delta \Phi}{\partial J_2} 
-\frac{\partial \delta f}{\partial J_1}
\frac{\partial \delta \Phi}{\partial w_1} - \frac{\partial \delta f}
{\partial J_2} \frac{\partial \delta \Phi}{\partial w_2} \,, 
\end{eqnarray}
if we use again the variables introduced in paper (I). $J_1$ denotes the
radial action integral of a stellar orbit and $J_2$ is an integral of motion
related to angular momentum. $w_1$ and $w_2$ are the canonical conjugate
variables, respectively. In expression (6) the perturbation of the distribution
function can be Fourier analyzed as
\begin{equation}
\delta f = \sum_{l_1} \int d l_2 f_{\rm {\bf l}} ({\bf J})
e^{i[l_1 w_1 + l_2 w_2]}\,.
\end{equation}
Since $w_1$ is a genuine angle variable, $l_1$ denotes integer numbers, whereas
$l_2$ is a real variable. The perturbation of the gravitational potential is
Fourier analyzed as
\begin{equation}
\delta \Phi = \int d k_{\rm x}\int d k_{\rm y} \Phi_{\rm {\bf k}} 
e^{i[ k_{\rm x} x + k_{\rm y} y]}
\end{equation}
where $x$ and $y$ denote the pseudo--Cartesian spatial coordinates of the
shearing sheet with $y$ pointing in the direction of rotation. Using the orbit
equations derived in paper (I) the Fourier expansion in equation (8) takes 
the form
\begin{eqnarray}
&& {\rm exp}\,i[k_{\rm x}x + k_{\rm y} y] =
{\rm exp}\,i \Big( k_{\rm x} \frac{J_2-\Omega_0}{-2B}  
\nonumber \\ && + k_{\rm x} 
\sqrt{\frac{2 J_1}{\kappa}}\sin{w_1} 
+ k_{\rm y} w_2 - k_{\rm y} \frac{\sqrt{2 \kappa J_1}}{2 B}\cos{w_1}\Big)
\end{eqnarray}
with Oort's second constant $B$ and the epicyclic frequency $\kappa$. We 
refer to paper (I) for the meaning of $\Omega_0$ and $r_0$, which define the 
center of the shearing sheet. Inserting the two Fourier components from 
equations (7) and (8) into the non--linear Poisson bracket (6) leads to
\begin{eqnarray}
&& [\delta f, \delta \Phi] =  \nonumber \\ &&
\Big( -l_1 \frac{\xi}{2J_1}\sin{(w_1-\overline{w})}
f_{\rm {\bf l}} - l_2 \frac{k_{\rm x}}{-2B} f_{\rm {\bf l}} \nonumber \\ &&
- i k_{\rm y} \frac{\partial f_{\rm {\bf l}}}{\partial J_2} - 
i\frac{\partial f_{\rm {\bf l}}}{\partial J_1}\xi \cos{(w_1-\overline{w})} \Big)
\Phi_{\rm {\bf k}} \nonumber \\ && \times {\rm exp}\,i\Big( l_1 w_1 + l_2 w_2
+ k_{\rm x} \frac{J_2-\Omega_0 r_0}{-2B} \nonumber \\ && + k_{\rm y} w_2
+ \xi\sin{(w_1-\overline{w})}\Big) \,,
\end{eqnarray}
where the auxiliary variables $\xi$ and $\overline{w}$ are defined as in paper
(I) as $\xi=\sqrt{\frac{2 J_1}{\kappa}} \sqrt{k_{\rm x}^2 + \frac{\kappa^2}{4
B^2} k_{\rm y}^2}$ and $\overline{w}=\arctan{(\frac{\kappa}{2B} 
\frac{k_{\rm y}}{k_{\rm x}})}$, respectively.

\subsection{Solution of the Boltzmann equation}

The Boltzmann equation (5) can be viewed as a linear partial differential
equation for  $\delta f$ with the two inhomogeneous terms $[f_0,\delta \Phi]$
and $[\delta f, \delta \Phi]$, respectively.
Formally it can be solved for each inhomogeneity separately and both 
solutions can be then combined as the final solution
\begin{equation}
\delta f =\delta f_1 + \delta f_2\,.
\end{equation}
The linearized Boltzmann equation 
\begin{equation}
\frac{\partial \delta f_1}{\partial t} + [\delta f_1,H_0] = 
- [f_0,\delta \Phi]
\end{equation}
has already been solved in paper (I) and we can use directly the result
obtained there. This leaves the second equation 
\begin{equation}
\frac{\partial \delta f_2}{\partial t} + [\delta f_2,H_0] = 
- [\delta f,\delta \Phi]
\end{equation}
to be treated here, which takes the explicit form 
\begin{eqnarray}
&&
\frac{\partial \delta f_2}{\partial t} + \kappa \frac{\partial \delta
f_2}{\partial w_{\rm{1}}} + \frac{A}{B} \left(J_{\rm{2}} - 
\Omega_{\rm{0}}r_{\rm{0}} \right) \frac{\partial \delta f_2}
{\partial w_{\rm{2}}} = \nonumber \\ &&
\Big( l_1 \frac{\xi}{2J_1}\sin{(w_1-\overline{w})}
f_{\rm {\bf l}} + l_2 \frac{k_{\rm x}}{-2B} f_{\rm {\bf l}} \nonumber \\ &&
+ i k_{\rm y} \frac{\partial f_{\rm {\bf l}}}{\partial J_2} + 
i\frac{\partial f_{\rm {\bf l}}}{\partial J_1}\xi \cos{(w_1-\overline{w})} \Big)
\Phi_{\rm {\bf k}} \nonumber \\ && \times {\rm exp}\,i\Big( l_1 w_1 + l_2 w_2
+ k_{\rm x} \frac{J_2-\Omega_0 r_0}{-2B} \nonumber \\ && + k_{\rm y} w_2
+ \xi\sin{(w_1-\overline{w})}\Big) 
\end{eqnarray}
with the first Oort constant $A=\Omega_0 + B$.
Equation (14) is of the same form as equation
(20) of paper (I) and can be treated in the same way. Since the
coefficients of the differential equation (14) do not depend on time $t$
or the variable $w_2$, equation (14) can be Fourier transformed with
respect to $t$ and $w_2$ leading to   
\begin{eqnarray}
&& i \omega f_{\rm{2},\omega} + \kappa \frac{d f_{\rm{2},\omega}}
{d w_{\rm{1}}} + i(k_{\rm{y}}+l_2) \frac{A}{B}\left( J_{\rm{2}}
-\Omega_{\rm{0}} r_{\rm{0}} \right) f_{\rm{2}, \omega} =\nonumber\\ &&
\Big( l_1 \frac{\xi}{2J_1}\sin{(w_1-\overline{w})}
f_{\rm {\bf l}}\Phi_{\rm {\bf k}}|_\omega
 + l_2 \frac{k_{\rm x}}{-2B} f_{\rm {\bf l}}\Phi_{\rm {\bf k}}|_\omega
  \nonumber \\ &&
+ i k_{\rm y} \frac{\partial f_{\rm {\bf l}}}{\partial J_2}
\Phi_{\rm {\bf k}}|_\omega  + 
i\frac{\partial f_{\rm {\bf l}}}{\partial J_1}\Phi_{\rm {\bf k}}|_\omega 
\xi \cos{(w_1-\overline{w})} \Big)
 \nonumber \\ && \times {\rm exp}\,i\Big( l_1 w_1 + l_2 w_2
+ k_{\rm x} \frac{J_2-\Omega_0 r_0}{-2B} \nonumber \\ && + k_{\rm y} w_2
+ \xi\sin{(w_1-\overline{w})}\Big) \,.
\end{eqnarray}
Notice that the products of $f_{\rm{\bf l}}$ and $\Phi_{\rm{\bf k}}$ and
$\frac{\partial f_{\rm{\bf l}}}{\partial J_{1,2}}$ and $\Phi_{\rm{\bf k}}$,
respectively,  have been Fourier transformed with respect to time
simultaneously,
which has to be taken into account when transforming back from frequency
to the time domain. Equation (15) is of exactly the same form as equation
(23) of paper (I) and can be solved in the same way by determining 
first the solution of the homogeneous part of the equation, then
constructing a particular integral of the inhomogeneous equation by
`variation of the constant', and finally adjusting the constant of
integration so that the solution is uniquely defined in velocity space.
We report here only the final result
\begin{eqnarray}
&& f_{\rm{2}, \omega}  = \frac{e^{i[\pi \eta +l_1 w_1]}}{2i\sin{\pi \eta}}
\exp{i[\omega t +(k_{\rm y} + l_2) w_2 + k_{\rm x}\frac{J_2 - \Omega_0 r_0}
{-2B}]} \nonumber \\ && \times
\Big[ \Big(\frac{1}{-2B} l_2 k_{\rm x} f_{\rm {\bf l}}\Phi_{\rm {\rm
k}}|_\omega + i k_{\rm y} \frac{\partial f_{\rm {\bf l}}}{\partial J_2}
\Phi_{\rm {\rm k}}|_\omega \Big) \nonumber \\ && \times
\int_{-2\pi}^0 dw_1' \exp{i [(l_1+\eta)w_1' +\xi \sin{(w_1'+w_1-\overline{w})}]}
\nonumber \\ &&
+l_1 \frac{\xi}{2J_1}f_{\rm {\bf l}}\Phi_{\rm {\rm k}}|_\omega
\int_{-2\pi}^0 dw_1' \sin{(w_1'+w_1-\overline{w})} \nonumber \\ && \times
\exp{i [(l_1+\eta)w_1' +\xi \sin{(w_1'+w_1-\overline{w})}]} \nonumber \\ &&
+i \xi \frac{\partial f_{\rm {\bf l}}}{\partial J_1}\Phi_{\rm {\rm k}}|_\omega
\int_{-2\pi}^0 dw_1' \cos{(w_1'+w_1-\overline{w})} \nonumber \\ && \times
\exp{i [(l_1+\eta)w_1' +\xi \sin{(w_1'+w_1-\overline{w})}]} \Big] \,,
\end{eqnarray}
where  $\eta = \frac{1}{\kappa}\Big(\omega + (k_{\rm{y}}+ l_2)
$ $\frac{A}{B}\left (J_{\rm{2}} - \Omega_{\rm{0}} r_{\rm{0}} \right)
\Big)$ has been introduced as an abbreviation to keep the formulae a little 
more compact. $w_1'$ is an auxiliary variable.

\subsection{Self consistent perturbations}

Even in the linear case the spatial pattern of the disk response is 
different to that of the imprinted potential perturbation
(cf.~equations 10 of paper I). In particular, through the dependence on 
$\eta$ and thus on the $J_{\rm{2}}$ variable there is a further dependence
of the disk response on the $x$ coordinate. This means that the
disk response to a Fourier component of the potential 
is {\em  not } a Fourier component of the general disk
response. Kalnajs (1971) has shown a way to overcome this difficulty by
integrating the disk response over all wave numbers and then Fourier
transforming this with respect to the spatial $x$ and $y$ coordinates
again. The result is integrated over velocity space in order to obtain the
surface density of the disk response and inserted into the Poisson equation
whose lhs is also Fourier transformed, 
\begin{eqnarray}
&&\frac{1}{\left(2\pi\right)^2}\int^{+\infty}_{-\infty} dz 
\int^{+\infty}_{-\infty} dx \int^{+\infty}_{-\infty} dy
e^{-i [k'_{\rm x} x + k'_{\rm y} y ]}\nonumber\\
&&\{\frac{\partial^{\rm{2}}}{\partial x^{\rm{2}}} +\frac{\partial^{\rm{2}}}
{\partial y^{\rm{2}}} +\frac{\partial^{\rm{2}}}{\partial z^{\rm{2}}} \}
\int^{+\infty}_{-\infty}dk_{\rm{x}}\int^{+\infty}_{-\infty}dk_{\rm{y}} 
\Phi_{\rm{\bf{k}}}|_\omega \nonumber \\
&&\times \exp\left[i [k_{\rm{x}} x + k_{\rm{y}} y] - 
\sqrt{k^{\rm{2}}_{\rm{x}} + k^{\rm{2}}_{\rm{y}}}\,|z|\right] \nonumber\\
&& = -2\sqrt{k^{\rm{2}}_{\rm{x}} + k^{\rm{2}}_{\rm{y}}}\, 
\Phi_{\rm{\bf{k}}}|_\omega \,
e^{ i [k_{\rm{x}} x + k_{\rm{y}} y ] }  \nonumber \\ &&
= \frac{4\pi G}{\left(2\pi\right)^{\rm{2}}} \int^{+\infty}_{-\infty}dx
\int^{+\infty}_{-\infty}dy e^{-i [k'_{\rm x} x + k'_{\rm y} y ]} \nonumber \\
&&\int^{+\infty}_{-\infty}dk_{\rm{x}} \int^{+\infty}_{-\infty}dk_{\rm{y}} 
\int^{+\infty}_{-\infty}du \int^{+\infty}_{-\infty}dv \delta f\,.
\end{eqnarray}
In this way self consistent
perturbations of a self gravitating disk can be calculated (cf.~equations 
35 and 36 of paper I). Accordingly, on the rhs of equation (17) a multiple 
integral over the disk response to the non--linear term of the form 
\begin{eqnarray}
&&\frac{4\pi G}{\left(2\pi\right)^{\rm{2}}} \int^{+\infty}_0 dJ_1
\int^{+\infty}_{-\infty} dJ_2 \int^{2 \pi}_0 dw_1 \int^{+\infty}_{-\infty}
dw_2 \nonumber \\ && 
\sum_{l_1 = -\infty}^\infty \int^{+\infty}_{-\infty} dl_2
\int^{+\infty}_{-\infty} dk_{\rm x} \int^{+\infty}_{-\infty} dk_{\rm y}
\nonumber \\ &&{\rm exp}-i\Big[k'_{\rm x}\frac{J_{\rm{2}} - 
\Omega_{\rm{0}} r_{\rm{0}}}{-2B} +k'_{\rm y} w_2 + \xi'\sin{(w_1-
\overline{w'})} \Big] \nonumber \\&& \times \frac{e^{i[\pi \eta +l_1 w_1]}}
{2i\sin{\pi \eta}}
\exp{i[\omega t +(k_{\rm y} + l_2) w_2 + k_{\rm x}\frac{J_2 - \Omega_0 r_0}
{-2B}]} \nonumber \\ && \times
\Big[ \Big(\frac{1}{-2B} l_2 k_{\rm x} f_{\rm {\bf l}}\Phi_{\rm {\rm
k}}|_\omega + i k_{\rm y} \frac{\partial f_{\rm {\bf l}}}{\partial J_2}
\Phi_{\rm {\rm k}}|_\omega \Big) \nonumber \\ && \times
\int_{-2\pi}^0 dw_1' \exp{i [(l_1+\eta)w_1' +\xi \sin{(w_1'+w_1-\overline{w})}]}
\nonumber \\ &&
+l_1 \frac{\xi}{2J_1}f_{\rm {\bf l}}\Phi_{\rm {\rm k}}|_\omega
\int_{-2\pi}^0 dw_1' \sin{(w_1'+w_1-\overline{w})} \nonumber \\ && \times
\exp{i [(l_1+\eta)w_1' +\xi \sin{(w_1'+w_1-\overline{w})}]} \nonumber \\ &&
+i \xi \frac{\partial f_{\rm {\bf l}}}{\partial J_1}\Phi_{\rm {\rm k}}|_\omega
\int_{-2\pi}^0 dw_1' \cos{(w_1'+w_1-\overline{w})} \nonumber \\ && \times
\exp{i [(l_1+\eta)w_1' +\xi \sin{(w_1'+w_1-\overline{w})}]} \Big]
\,.
\end{eqnarray}
has to be considered. Again we have changed from spatial 
($x$, $y$) and ($u$, $v$) velocity coordinates to ($J_1$, $J_2$) and 
($w_1$, $w_2$) coordinates. $\xi'$ and and $\overline{w'}$ are defined in an
analogous way to $\xi$ and $\overline{w}$, respectively, for primed wave
numbers. The integrations with respect to $J_1$ and $J_2$ cannot be carried out
directly, because $f_{\rm {\bf l}}({\bf J})$ is not known explicitely. 
Integrations over the angle variables $w_1$ and $w'_1$ lead to intermediate 
results which are not enlightening. But the integration over the $w_{\rm{2}}$ 
variable gives immediately
\begin{equation}
\frac{1}{2\pi}\int^{+\infty}_{-\infty}dw_{\rm{2}} e^{i[ k_{\rm y} 
+l_2 - k'_{\rm y} ] w_2} = 
\delta\left(k_{\rm{y}} + l_2 - k'_{\rm{y}}\right)\,.
\end{equation}
The integration over the wave number $k_{\rm{y}}$ contracts the integrand of 
(18) to $k_{\rm{y}} = k'_{\rm{y}}-l_2$, which is the customary sum rule of wave
numbers well known from plasma physics. There is an analogous sum rule for the
$k_{\rm x}$ wave numbers, which is, however, not directly obvious from
expression (18) because of the different Fourier transforms in equations (7)
and (8).

\subsection{Volterra integral equation} 

It was shown in paper (I) that the solution of the linearized Boltzmann 
equation (12) leads, upon inserting it into the Fourier transformed Poisson 
equation (17), to the Volterra integral equation ($k'_{\rm y} > 0$)
\begin{equation}
\Phi_{\rm {\bf k'}} = \int_{-\infty}^{k'_{\rm{x}}} dk_{\rm{x}} \mathcal{K} 
\left(k_{\rm{x}},k'_{\rm{x}}\right)\Phi_{\rm{\bf{k}}} + r^l_{\rm{\bf{k'}}}\,,
\end{equation}
where the kernel $\mathcal{K}$ is given by equation (67) of paper (I).  
$r^l_{\rm{\bf{k'}}}$ describes an inhomogeneity of equation (20) related either 
to an initial non--equilibrium state of the shearing sheet or an external
perturbation. The formal solution of equation (13) as derived in the previous 
section and then inserted into the Fourier transformed Poisson equation (17) has
to be added to the rhs of equation (20) in order to obtain the solution of the
non--linear Boltzmann equation. This means formally a further inhomogeneity 
of the Volterra equation (20), $r^{nl}_{\rm{\bf{k'}}}$, which is given by
(cf.~equation 36 of paper I)
\begin{eqnarray}
 && r_{\rm {\bf k'}}^{nl} = \frac{2G}{-2\sqrt{k{'_{\rm x}}^2 + k{'_{\rm y}}^2}}
\int^{+\infty}_0 dJ_1
\int^{+\infty}_{-\infty} dJ_2 \int^{2 \pi}_0 dw_1 \nonumber \\ && 
\sum_{l_1 = -\infty}^\infty \int^{+\infty}_{-\infty} dl_2
\int^{+\infty}_{-\infty} dk_{\rm x} 
\nonumber \\ &&{\rm exp}-i\Big[k'_{\rm x}\frac{J_{\rm{2}} - 
\Omega_{\rm{0}} r_{\rm{0}}}{-2B} + \xi'\sin{(w_1-
\overline{w'})} \Big] \nonumber \\&& \times \frac{e^{i[\pi \eta +l_1 w_1]}}
{2i\sin{\pi \eta}}
\exp{i[\omega t +k'_{\rm y} w_2 + k_{\rm x}\frac{J_2 - \Omega_0 r_0}
{-2B}]} \nonumber \\ && \times
\Big[ \Big(\frac{1}{-2B} l_2 k_{\rm x} f_{\rm {\bf l}}\Phi_{\rm {\rm
k}}|_\omega + i k_{\rm y} \frac{\partial f_{\rm {\bf l}}}{\partial J_2}
\Phi_{\rm {\rm k}}|_\omega \Big) \nonumber \\ && \times
\int_{-2\pi}^0 dw_1' \exp{i [(l_1+\eta)w_1' +\xi \sin{(w_1'+w_1-\overline{w})}]}
\nonumber \\ &&
+l_1 \frac{\xi}{2J_1}f_{\rm {\bf l}}\Phi_{\rm {\rm k}}|_\omega
\int_{-2\pi}^0 dw_1' \sin{(w_1'+w_1-\overline{w})} \nonumber \\ && \times
\exp{i [(l_1+\eta)w_1' +\xi \sin{(w_1'+w_1-\overline{w})}]} \nonumber \\ &&
+i \xi \frac{\partial f_{\rm {\bf l}}}{\partial J_1}\Phi_{\rm {\rm k}}|_\omega
\int_{-2\pi}^0 dw_1' \cos{(w_1'+w_1-\overline{w})} \nonumber \\ && \times
\exp{i [(l_1+\eta)w_1' +\xi \sin{(w_1'+w_1-\overline{w})}]} \Big]
\,. 
\end{eqnarray}
The Volterra equation (20) is derived in the frequency domain. If transformed
back to the time domain, we find for the linear part
\begin{eqnarray}
&&  \Phi_{\rm{\bf k'}}(t) = 2 A |k'_{\rm y}| \int_{-\infty}^{t} dt' 
\widetilde{\mathcal{K}}\left(k'_{\rm{x}}- 2 A k'_{\rm y}(t-t'),k'_{\rm{x}}
\right)\nonumber \\ && \times
 \Phi_{\rm k'_{\rm x} - 2 A k'_{\rm y}(t-t'),k'_{\rm y}}(t') 
+ r_{\rm {\bf k'}}^l(t) \,,
\end{eqnarray}
where the overhead tilde indicates that the $\omega$--dependent term of the 
kernel $\mathcal{K}$ has been split off. In deriving equation (22), use of the 
convolution theorem of the Fourier transform of products of functions has 
been made (cf.~paper I). In order to transform the non--linear term back into 
the time domain we consider first in expression (21) the terms that depend 
explicitely on $\omega$, 
\begin{eqnarray}
&&\int_{-\infty +i\nu}^{\infty +i\nu} d\omega e^{i[\omega t +\pi \eta
+\eta w'_1]} \frac{1}{\sin{(\pi \eta)}} = \\ &&
\kappa \int_{-\infty +i\frac{\nu}{\kappa}}^{\infty +i\frac{\nu}{\kappa}} d\eta 
e^{i[(\kappa \eta -\frac{A}{B}(J_2 -\Omega_0r_0)k'_{\rm y})t + \eta w'_1]}
\frac{e^{i\pi \eta}}{\sin{(\pi \eta)}} \nonumber
\end{eqnarray}
with $\eta = \frac{1}{\kappa}(\omega + \frac{A}{B}(J_2-\Omega_0 r_0)k'_{\rm
y})$. According to Landau's rule the imaginary part of the frequency is chosen
negative, $\Im (\omega) = \nu < 0$. By deforming the integration contour in the
complex frequency plane we find that expression (23) is proportional to the
unit step function
\begin{equation}
u(\kappa t +w'_1)=\left\{ \begin{array}{cc} 1, & \kappa t +w'_1 \geq 0 \\
 0, & \kappa t +w'_1 < 0 \end{array} \right.\,.
\end{equation}
Thus the inhomogeneity (21) of the Volterra equation can be cast formally into
the form
\begin{eqnarray}
&&  r_{\rm {\bf k'}}^{nl}(t) = 2G \int^{+\infty}_0 dJ_1
\int^{+\infty}_{-\infty} dJ_2 \int^{2 \pi}_0 dw_1 \nonumber \\ && 
\sum_{l_1 = -\infty}^\infty \int^{+\infty}_{-\infty} dl_2
\int^{+\infty}_{-\infty} dk_{\rm x} \int^0_{-2 \pi} dw'_1 
\int_{-\infty}^{t+w'_1/\kappa} dt' \nonumber \\ &&
\Big[ \mathcal{N}_1({\bf J},w_1,w'_1,{\bf
l},k_{\rm x},{\bf k'},t-t') f_{\rm {\bf l}}(t')\Phi_{\rm k_{\rm x},k'_{\rm
y}-l_2}(t') \nonumber \\ && +
\mathcal{N}_2({\bf J},w_1,w'_1,{\bf l},k_{\rm x},{\bf k'},t-t') 
\frac{\partial f_{\rm {\bf l}}}{\partial J_1}|_{t'}\Phi_{\rm k_{\rm x},k'_{\rm
y}-l_2}(t') \nonumber \\ && +
\mathcal{N}_3({\bf J},w_1,w'_1,{\bf l},k_{\rm x},{\bf k'},t-t') 
\frac{\partial f_{\rm {\bf l}}}{\partial J_2}|_{t'}\Phi_{\rm k_{\rm x},k'_{\rm
y}-l_2}(t')\Big]\,,
\end{eqnarray}
which has to be added to the rhs of equation (22). Again use of the 
convolution theorem of the Fourier transform of products of functions has
been made. Since $w'_1$ is less than or equal to zero, the integration over
time $t'$ extends to the upper limit $t$.

\subsection{Discussion of the Volterra equation}

The Volterra equation (22) augmented by $r_{\rm {\bf k'}}^{nl}$ reveals
clearly the role of the non--linear term of the Boltzmann equation in the
dynamical evolution of the shearing sheet.

In Fig.~1 the linear evolution of a density wave, calculated neglecting the 
non--linear part of equation (22), is illustrated. The density wave is 
initialized as
\begin{equation}
r_{\rm{\bf k'}} \propto  \delta(k'_{\rm x}-
(k_{\rm x}^{\rm in}+2 A k'_{\rm y} t))
\delta(k'_{\rm y}-k_{\rm y}^{\rm in}) e^{-\frac{t^2}{T^2}}\,,
\end{equation}
where $T$ is a time constant.
As can be seen from Fig.~1 or directly from the interplay of the indices of
the Fourier coefficients of the gravitational potential in the linear
equation (22), $k'_{\rm x}-2 A k'_{\rm y}(t-t')$, with  
time $t'$ at which the Fourier coefficients are taken, the delta impulse
travels at constant wave number $k_{\rm y}^{\rm in}$ to positive wave numbers
$k_{\rm x}$ at a speed of $\dot{k}_{\rm x}^{\rm eff} = 
2 A k_{\rm y}^{\rm in}$. The wave crests of the 
density waves are oriented perpendicular to the wave vectors, which means 
that the wave crests swing around from leading to trailing orientation. While
swinging around the wave is amplified and then dies out. In addition the
amplitude shows oscillatory behaviour. Trailing density waves are
not effectively amplified (cf.~paper I). Since no outer boundaries
have been introduced, no rigidly rotating spiral modes appear in the shearing 
sheet (cf.~paper II).
\begin{figure}
\resizebox{\hsize}{!}{\includegraphics {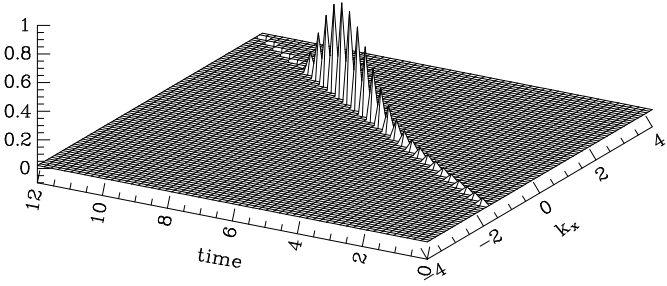}}
\caption{Linear swing amplification of a density wave in the shearing sheet.
The wave is initialized as a delta like impulse in wave number space with an
initial wave vector $(k_{\rm x}^{\rm in}, k_{\rm y}^{\rm in})=
(-2,0.5)\,k_{\rm crit}$, which corresponds to initially leading arms, and an 
initial amplitude of 0.04. The impulse travels at constant
wave number $k_{\rm y}^{\rm in}$ with an effective radial wave number 
$k_{\rm x}^{\rm eff}= k_{\rm x}^{\rm in} +2 A k_{\rm y}^{\rm in} t$. The wave
numbers are given in units of $k_{\rm crit}$ defined below,
and the time unit is $\Omega_0^{-1}$, the inverse of the mean angular 
velocity of the shearing sheet. Negative amplitudes at $t > 9.5 
\Omega_0^{-1}$ are not shown. The parameters of the disk model are
$A/\Omega_0$ = 0.5 and $Q$ = 1.4.}
\label{fig:1}       
\end{figure}
If the non--linear term is taken into account in equation (22), the evolution of
the shearing sheet becomes more complicated. In the non--linear term (25)
Fourier coefficients of the perturbations of the distribution function and the
gravitational potential are combined. Since the non--linear term represents an
inhomogeneity of the Volterra integral equation, each combination of the Fourier
coefficients sampled over times $t'$ earlier than $t$ provides fresh input for
the swing amplification mechanism. It is instructive to rewrite the
inhomogeneity $r_{\rm {\bf k'}}^{nl}$ given by equation (25) formally as
\begin{equation}
r_{\rm {\bf k'}}^{nl}(t) = \int_{-\infty}^t dt_0 \delta(t-t_0)
r_{\rm {\bf k'}}^{nl}(t_0)
\end{equation} 
and to consider for each $t_0$--epoch the corresponding Volterra equation
\begin{eqnarray}
&& \Phi_{\rm{\bf{k'},t_0}} (t)= 2 A |k'_{\rm y}| \int_{-\infty}^t 
dt' \widetilde{\mathcal{K}} 
\left(k'_{\rm{x}}-2 A k'_{\rm y}(t-t'),k'_{\rm{x}}\right) \nonumber \\ && 
\times \Phi_{\rm k'_{\rm x} -2 A k'_{\rm y}(t-t'), k'_{\rm y},t_0}(t') + 
\delta(t-t_0) r_{\rm {\bf k'}}^{nl}(t_0)\,.
\end{eqnarray}
Obviously the full Volterra equation is given by the superposition 
\begin{eqnarray}
&& \Phi_{\rm {\bf k'}}(t) = \int_{-\infty}^t dt_0 \Phi_{\rm {\bf k'}, t_0}(t) =
\nonumber \\ && 2 A |k'_{\rm y}| \int_{-\infty}^t 
dt' \widetilde{\mathcal{K}} 
\left(k'_{\rm{x}}-2 A k'_{\rm y}(t-t'),k'_{\rm{x}}\right) \nonumber \\ && 
\times \int_{-\infty}^t dt_0
\Phi_{\rm k'_{\rm x} -2 A k'_{\rm y}(t-t'), k'_{\rm y},t_0}(t') + 
r_{\rm {\bf k'}}^{nl}(t)\,.
\end{eqnarray}
As can be seen from equation (28) each $\delta$--impulse, sampling the 
history of the evolution of the sheet prior to $t_0$, is propagated linearly to
time $t$ {\em independently} of the following evolution of the sheet. On the
other hand, these linearly evolving $\delta$--impulses influence later
$\delta$--impulses by non--linear interaction of the Fourier components, always
cascading off fresh input for the swing amplification mechanism. Only at $t_0$ =
$t$ is there an instantaneous non--linear interaction of the Fourier components.
Thus the non--linear interactions lead {\em inter alia} to a feedback cycle for 
swing amplification. Even the interaction of a pair of trailing Fourier
components (${\bf k} > 0$), which are decaying, if coupled so that 
their wave vectors are subtracted from each other, leads to leading density
waves as an input to the swing amplification mechanism, which is then amplified
accordingly. Combined with some residual particle
noise this might well account for the recurrent spiral instabilities reported 
from the numerical simulations. As mentioned above, Sellwood \& Carlberg (1984) 
have demonstrated by analyzing the evolution of individual spiral arms in their
simulations that these are indeed swing amplified density waves of the same 
type as in the shearing sheet.
The effect of the non--linear term of the Boltzmann equation thus has aspects
that are completely different to the effects of non--linear terms in the 
hydrodynamical or Jeans equations. If the latter are used to describe the 
dynamics of density waves, the non--linear advective and enthalpy terms imply
instantaneous non--linear interaction of the Fourier components and have been
shown to lead to saturation and soliton--like self--modulation effects of the
density waves (Ikeuchi \& Nakamura 1976, Norman 1978, Meinel 1983).

The feedback cycle described here is of course an effect depending 
quadratically on the magnitude of the perturbations of the sheet. Thus,
if the amplitudes of the density waves are at very low levels, the efficiency 
of the swing amplification mechanism might not be sufficient to sustain the 
cycle and it will die out eventually.

\section{Numerical self consistent field simulations}

In the previous section we have developed the theoretical concept of a feedback
cycle for density waves in the shearing sheet mitigated by the non--linear
coupling of the waves. In order to demonstrate concrete examples of this
effect we have run numerical simulations of the dynamical evolution of the
shearing sheet.

\subsection{Self consistent field method (SCF)}

The SCF method relies on a complete set of pairwise basis functions into which 
the density of the particles and the gravitational potential are simultaneously
expanded. Each of the pairwise basis functions solve the Poisson 
equation. For the shearing sheet these are simply Fourier transforms,
\begin{equation}
\Sigma({\bf R}) = \int d^2k \Sigma_{\rm {\bf k}} e^{i\,({\bf k},{\bf R})}\,,
\end{equation}
for the surface density of the sheet, and 
\begin{equation}
\Phi({\bf R},z=0) = \int d^2k \Phi_{\rm {\bf k}} e^{i\,({\bf k},{\bf R})}
\end{equation}
with
\begin{equation}
\Phi_{\rm {\bf k}} = - \frac{2 \pi G}{k} \Sigma_{\rm {\bf k}}
\end{equation}
for the gravitational potential, where $k = |{\bf k}|$ (cf.~equations 32 to 34 
of paper I). Changing to real quantities the transformation pair can be 
written as
\begin{eqnarray}
 & & A({\bf k}) = \int d^2R\,\Sigma({\rm {\bf R}}) \cos{({\bf k},{\bf R})}\,,
 \quad A({\bf -k})=A({\bf k})\,, \nonumber \\ & & 
B({\bf k}) = \int d^2R\,\Sigma({\rm {\bf R}}) \sin{({\bf k},{\bf R})}\,, \quad
B({\bf -k})=-B({\bf k})\,, \nonumber \\
 & & \Sigma({\bf R}) = \frac{1}{2 \pi^2} \int_{-\infty}^\infty dk_{\rm x}
 \int_0^\infty dk_{\rm y}(A({\bf k})\cos{({\bf k},{\bf R})} \nonumber \\ && +
 B({\bf k})\sin{({\bf k},{\bf R})})\,, 
  \nonumber \\ & & 
\Phi({\bf R},z=0) = -\frac{G}{\pi}\int_{-\infty}^\infty dk_{\rm x}
 \int_0^\infty dk_{\rm y} \nonumber \\ && \times
 \frac{1}{k}(A({\bf k})\cos{({\bf k},{\bf R})}
 + B({\bf k})\sin{({\bf k},{\bf R})}) \,.
\end{eqnarray}
In the simulation the surface density is given by
\begin{equation}
\Sigma({\bf R}) = \sum_{\rm j = 1}^N \Delta \Sigma\,\delta (x-x_{\rm j}) \delta
(y-y_{\rm j})\,,
\end{equation}
where $(x_{\rm j}, y_{\rm j})$ denotes the position of particle $j$. 
$\Delta \Sigma$ is its contribution to the surface density, $\Delta \Sigma =
\Sigma_0/N$. Accordingly we find from equations (33)
\begin{eqnarray}
 && A({\bf k}) = \sum_{\rm j = 1}^N \Delta \Sigma\,
 \cos{({\bf k},{\bf R_{\rm j}})} 
 - 4 \Sigma_0 \frac{\sin{(k_{\rm x} x_0)}}{k_{\rm x}}
 \frac{\sin{(k_{\rm y} y_0)}}{k_{\rm y}} \,,\nonumber \\ &&
B({\bf k}) = \sum_{\rm j = 1}^N \Delta \Sigma\,
 \sin{({\bf k},{\bf R_{\rm j}})} \,. 
\end{eqnarray}
In practice we cannot simulate an infinite sheet, but choose a finite size $2x_0
\times 2y_0$. This active sheet is surrounded by virtual sheets of the same size
sliding along the central sheet according to the linear shear flow. If particles
are leaving the active sheet across one side, they are re--fed into the opposite
side as if entering from one of the neighbouring virtual sheets with exactly 
the same particle distribution as in the active sheet (Wisdom \& Tremaine 1988,
Toomre 1990). Moreover, the SCF method is meant to simulate the fluctuations of
the surface density of the sheet. Thus we have subtracted in equations (35)
the Fourier coefficients of the unperturbed background density $\Sigma_0$
\begin{eqnarray}
&& \int_{\rm -x_0}^{\rm x_0} dx \int_{\rm -y_0}^{\rm y_0} dy\,
\Sigma_0\, e^{-i({\bf k},{\bf R})} = \nonumber \\ &&
  4 \Sigma_0 \frac{\sin{(k_{\rm x} x_0)}}{k_{\rm x}}
 \frac{\sin{(k_{\rm y} y_0)}}{k_{\rm y}} \,.
\end{eqnarray}
Otherwise there would be an extra force pulling all particles towards the center
of the sheet.

We have discussed in the introduction that the reason for the success of the 
infinite shearing sheet model describing spiral density waves realistically is 
the rapid convergence of the Poisson integral in self gravitating disks and that
the ``effective range'' of gravity of the
density waves is only of the order $ \lambda_{\rm crit}/2$. Thus, if the size of
the simulated sheet is chosen large enough, it should behave like an infinite
shearing sheet.
\begin{figure}
\centering
%
%
\caption[]{Initial set up of the 32768 particles in the simulation of the 
shearing sheet. The $y$--axis points into the direction of the shear flow.
The size is $2 x_0 \times 2 y_0 = 20 \times 28 \lambda_{\rm crit}$.}
\label{fig:2a}       
\end{figure}
Next the wave numbers are discretized with bin widths $\Delta k_{\rm x}$ and
$\Delta k_{\rm y}$, respectively. Discretizing equations (33) accordingly we
obtain for the planar components of the gradient of the gravitational potential 
\begin{eqnarray}
&& -\nabla \Phi ({\bf R}_{\rm j}, z=0) = \sum_{\rm k_{\rm x,n}}\Delta
{\rm k_{\rm x}}\sum_{\rm k_{\rm y,n}}\Delta {\rm k_{\rm y}}
\frac{G {\bf k}_{\rm n}}{\pi \sqrt{k_{\rm x,n}^2+ k_{\rm y,n}^2}} \nonumber
\\ && \times [-A({\bf k}_{\rm n}) \sin{({\bf k}_{\rm n},{\bf R}_{\rm j})}
+B({\bf k}_{\rm n}) \cos{({\bf k}_{\rm n},{\bf R}_{\rm j})}]\,,
\end{eqnarray}
with ${\bf k}_{\rm n}= (k_{\rm x,n},k_{\rm y,n})$. In each time step of the
numerical algorithm the coefficients $A({\bf k}_{\rm n})$ and 
$B({\bf k}_{\rm n})$ and the forces (37) due to the fluctuations in the sheet
are calculated anew. Then all particles are moved one time step ahead. The 
equations of motion are the epicyclic equations of motion augmented by the 
fluctuating forces (Toomre 1981, 1990),
\begin{eqnarray}
&& \ddot{x} = 2 \Omega_0 \dot{y} + 4 \Omega_0 A - \frac{\partial
\Phi}{\partial x}(x,y)\,,\quad u = \dot{x}\,, \nonumber \\ &&
\ddot{y} = -2 \Omega_0 \dot{x} - \frac{\partial\Phi}{\partial y}(x,y)\,, 
\quad v = \dot{y}+2 A x \,.
\end{eqnarray}
In a quiet disk the fluctuating forces would be zero and the 
particles would follow simply epicyclic orbits. In the simulations the 
equations of motion (38) are integrated numerically using the symplectic 
integrator described in the appendix.

\subsection{Trial runs}

The basic dynamics of the shearing sheet is characterized by Oort's constants
$A=-\frac{1}{2} R_0 \frac{d \Omega_0}{d R_0}$, $B=A-\Omega_0$, the epicyclic
frequency of the particle orbits $\kappa=\sqrt{-4 \Omega_0 B }$, the
epicyclic ratio of the velocity dispersions of the particles
$\sigma_u^2/\sigma_v^2=4 \Omega_0^2/\kappa^2$, the critical wave number
$k_{\rm crit}=\kappa^2/(2 \pi G \Sigma_0)$ and the critical wave length
$\lambda_{\rm crit}=2 \pi/k_{\rm crit}$, and finally the Toomre stability
parameter $Q=\kappa \sigma_u/(3.36 G \Sigma_0)$. We introduce
dimensionless model units in the following way. The length unit is 
$\lambda_{\rm crit} =1$ and accordingly the wave number unit 
$k_{\rm crit}=2 \pi$. We set $G \Sigma_0=1$ which implies $\kappa=2 \pi$. The
time unit is then the epicyclic period $t_\kappa = 1$. Moreover we assume a flat
rotation curve, $A=-B=\frac{1}{2} \Omega_0$ so that
$A=-B=\frac{\pi}{\sqrt{2}}$ and $\Omega_0= \sqrt{2}\pi$. The radial velocity
dispersion is given by $\sigma_u=\frac{3.36}{2\pi}Q$ and the epicyclic ratio
is $\sigma_u/\sigma_v=\sqrt{2}$. In all the simulations presented here we have 
used the following parameters:\\
number of particles $N$ = 32768\\
size $x_0$=10, $y_0$ = $\sqrt{2} x_0$ = 14.1\\
time step 0.03125\\
bin size of wave numbers $\Delta k{\rm x}=\Delta k{\rm y}$ = 0.0625\\
64 bins in $k_{\rm x}$ ranging from -4 to 4\\
32 bins in $k_{\rm y}$ ranging from 0 to 4\\
$Q$=1.2\\

The simulations were set up as illustrated in Fig.~2 according to Sellwood's
(1983) quiet start recipe where the particles are distributed spatially
homogeneously and according to a Schwarzschild distribution in velocity space.
\begin{figure*}
\centering
%
%
\caption[]{Dynamical evolution of the shearing sheet. Snapshots of particle
positions are shown for $t$ = 0.5, 1, ... 3.5, and 4, respectively. 
The $y$--axis is oriented in the direction of the shear flow. The size of each
frame is the same as in Fig.~2}
\label{fig:2b}       
\end{figure*}
\begin{figure}
\resizebox{\hsize}{!}{\includegraphics {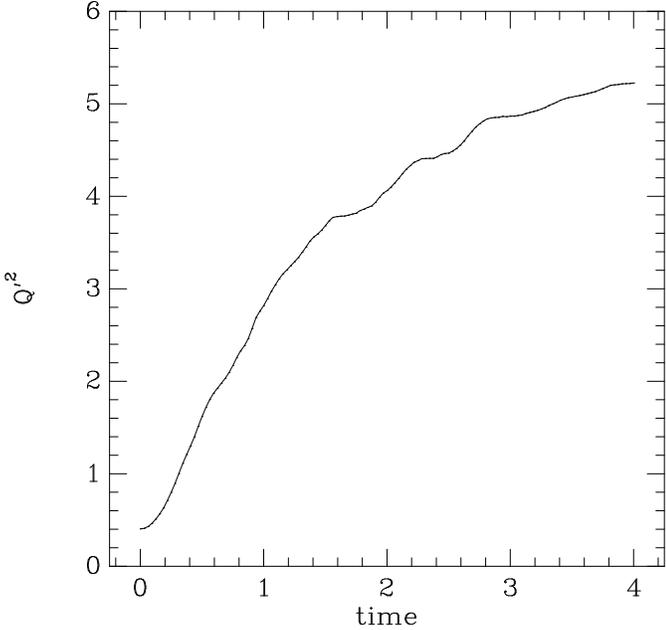}}
\caption{The rise of $Q'^2 = \langle u^2 \rangle /\Sigma_0^2$ as function of 
time illustrating the disk heating. The time unit is the epicyclic period.}
\label{fig:3}       
\end{figure}
\begin{figure*}
\centering
%
%
\caption[]{Same as Fig.~3, but the shearing sheet is dynamically cooled by
accretion of particles during the simulation.}
\label{fig:4}       
\end{figure*}
In Fig.~3 we show the simulation of the dynamical evolution of the shearing
sheet over 4 time units as snapshots of the 32768 particle positions. The sheet,
which was initially homogeneous, quickly develops density waves which can be 
seen as streaks in the frames in Fig.~3 in the time interval 
$t \approx$ 1 to 2. However, the sheet then becomes featureless again. This is 
due to rapid dynamical heating of the sheet which was already observed by
Sellwood \& Carlberg (1984) and in other numerical simulations
of the dynamical evolution of galactic disks such as by Tomley et al.~(1991), 
Toomre (1990), Toomre \& Kalnajs (1991), Chiueh \& Tseng (2000).
We illustrate the dynamical heating in Fig.~4 where we show the evolution of 
the ratio $Q'^2=\langle u^2 \rangle /\Sigma_0^2$, which mimics the $Q$ 
parameter, as function of time. During the evolution of the sheet its value has 
risen from 0.4 to 5.2, which reflects the rise of the velocity dispersion of 
the particles with time and that the sheet has become dynamically hot. The 
nearly linear rise of $Q'^2$ in the early phases of the simulation nicely fits 
the dynamical disk heating effect of density waves described 
theoretically by Fuchs (2001b) and Griv et al.~(2002).
In order to remedy this effect we have cooled the sheet dynamically like
Sellwood \& Carlberg (1984) by adding during the simulation particles on orbits
with their initial epicycle amplitudes. In Figs.~5 and 
6 we show the results of a simulation which grew linearly from 16384 particles 
to 32768 particles as in the previous simulation. As can be seen from Figs.~5 
and 6 the density wave activity and the pseudo--$Q$ parameter stayed fairly 
constant throughout the simulation, even though the velocity dispersion of the
particles was rising. Since the $Q'$ parameter is larger than in the 
initial stages of the previous run, the amplitudes of the density waves shown 
in Fig.~5 are smaller than in the early frames of Fig.~3.
\begin{figure}
\resizebox{\hsize}{!}{\includegraphics {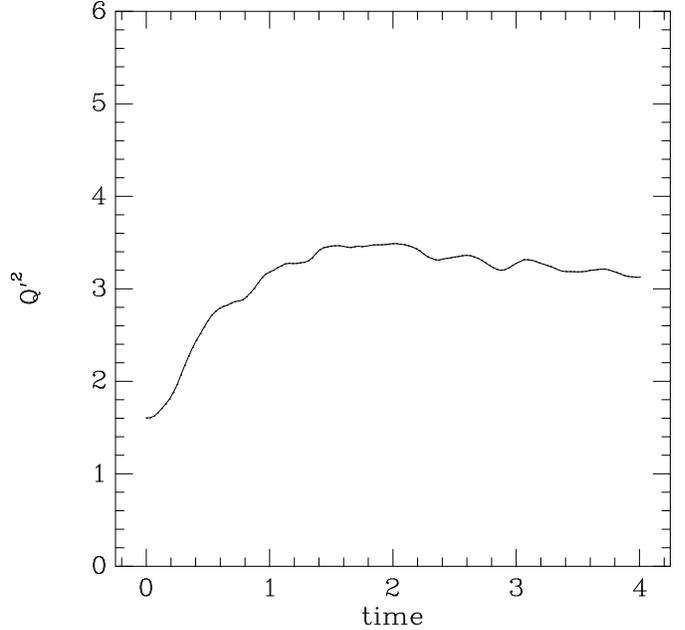}}
\caption{Same as Fig.~4, but the shearing sheet is dynamically cooled by
accretion of particles during the simulation.}
\label{fig:5}       
\end{figure}
We illustrate the tips of the spiked amplitude distribution of the $A({\bf k})$
Fourier coefficients of the density waves in Fig.~7. The distribution of the
$B({\bf k})$ Fourier coefficients is very similar. As expected from linear 
theory (Julian \& Toomre 1965, Toomre 1981, Fuchs 2001a) density waves with
wave numbers in the region around ${\bf k} \approx (1, 0.5)\,k_{\rm crit}$
implying pitch angles of about $27^\circ$ have 
the largest amplitudes. These are the streaks seen in Fig.~5. They are, however,
not stationary patterns, but formed by the superposition of transient shearing 
density waves. A cut through the power spectrum 
$\sqrt{A^2({\bf k})+B^2({\bf k})}$ of the density wave amplitudes at a given 
$k_{\rm y}$ wave number is also instructive. In Fig.~8 cuts at
$k_{\rm y}=0.5\,k_{\rm crit}$ are
illustrated for the initial ($t=0$) and a later stage ($t=2$) of the simulation.
Initially the power spectrum is flat in accordance with random distribution of
the disk particles. Later on  the power spectrum develops not only the high peak
at positive $k_{\rm x} \approx 1\,k_{\rm crit}$ wave numbers which corresponds 
to the maximally swing amplified amplitudes of the shearing density waves, but 
is also significantly higher at negative $k_{\rm x}$ wave numbers than the power 
spectrum of randomly distributed particles. Such leading density waves are the
input to the swing amplification mechanism. We conclude in agreement with
Sellwood \& Carlberg (1984), who presented a similar argument, that the
recurring density waves are indeed not amplified random noise, but are
connected events. The enhanced power at leading wave numbers supports the 
argument that the coupling of trailing density waves can lead to leading input 
to the swing amplifier.

The SCF simulations shown in Fig.~5 look very similar to the
direct N--body simulations of the dynamical evolution of the shearing sheet by
Toomre (1990) and Toomre \& Kalnajs (1991). In these simulations an artificial
friction force acting on the particles in the sheet is included in order to
overcome the disk heating problem. Toomre \& Kalnajs (1991) explain the 
`hotch--potch of swirling density waves' as the collective response of the 
shearing sheet to each of its individual members by polarization
clouds around each particle. Instead of this quasi--linear concept we prefer to
interpret our simulations in terms of the feedback cycle described in section
(2). The SCF simulations illustrated in Fig.~5 resemble closely 
the direct N--body simulations of Chiueh \& Tseng (2000, their run a). Disk 
heating effects are apparently controlled there by the choice the softening 
length.
\begin{figure}
\centering
\resizebox{\hsize}{!}{\includegraphics {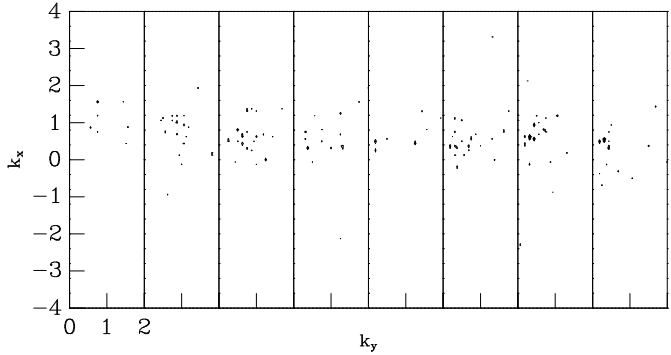}}
\caption{Peaks of positive $A({\bf k})$ Fourier coefficients in the simulations 
presented in Fig.~5 at times $t$ = 0.5, 1, ... 3.5, and 4, respectively. 
For clarity only contour levels of at least 40\% of the maxima of the
spikes are shown. The wave numbers are given in units of $k_{\rm crit}$.}
\label{fig:7}       
\end{figure}
\begin{figure}
\centering
\resizebox{\hsize}{!}{\includegraphics {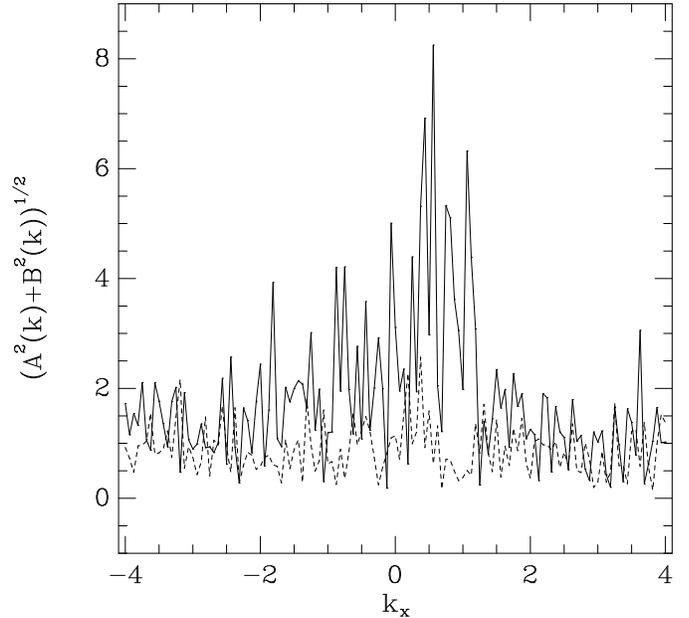}}
\caption{Power spectrum $\sqrt{A^2(k_{\rm x},0.5)+B^2(k_{\rm x},0.5)}$ of the
density wave amplitudes in the simulation shown in Fig.~5. The dashed and solid
lines correspond to times $t=0$ and $t=2$, respectively. The wave numbers are 
given in units of $k_{\rm crit}$.}
\label{fig:8}       
\end{figure}

\subsection{Exciting a swing amplified density wave}

The numerically simulated shearing sheet is subjected to a ``strong'' external
potential perturbation of the form considered in section (2.5),
\begin{equation}
\delta \Phi_{\rm{\bf k}} \propto \delta(k_{\rm x}-(k_{\rm x}^{in}+
2 A k_{\rm y}t)) \delta(k_{\rm y}-k_{\rm y}^{in})e^{-\frac{t^2}{T^2}}\,.
\end{equation}
It was shown in paper (I) that according to
linear theory the shearing sheet responds with a swing amplified shearing
density wave. In Fig.~9 the result of a simulation with a time 
constant of $T$ = 1.8 is traced in ${\bf k}$--wave--vector space. As can
be seen from Fig.~9 the disk response is indeed of exactly the same nature as 
predicted by linear theory. The impulse travels from negative $k_{\rm x}$ to
positive wave numbers and the amplitude is maximally amplified around
$k_{\rm x} \approx 2$.

\subsection{Non--linear wave coupling}

\begin{figure*}
\centering
%
%
\caption[]{Response of the shearing sheet to an external impulsive potential
perturbation with an initial wave vector $(k_{\rm x}, k_{\rm y}) = 
(-2, 0.5)\,k_{\rm crit}$ traced in wave number space. Frames are shown for time 
$t$ = 0, 0.25,..., 1.5, and 1.75, respectively.}
\label{fig:9}       
\end{figure*}
\begin{figure*}
\centering
%
%
\caption[]{Response of the shearing sheet to two external impulsive potential
perturbations with initial wave vectors $(k_{\rm x}, k_{\rm y}) = 
(-2, 0.5)\,k_{\rm crit}$ and $(k_{\rm x}, k_{\rm y})= (-3, 1)\,k_{\rm crit}$,
respectively traced in wave number space. Frames are shown for time $t$ = 
0, 0.25, 0.5, 0.88, 1.0, 1.13, 1.25, and 1.38, respectively.}
\label{fig:10}       
\end{figure*}
\begin{figure*}
\centering
%
%
\caption{Same as Fig.~8, but with two external impulsive potential
perturbations with initial wave vectors $(k_{\rm x}, k_{\rm y})$ = (-2, 0.5)
$k_{\rm crit}$ and $(k_{\rm x}, k_{\rm y}) = (-3, 1.5)\,k_{\rm crit}$,
respectively. Frames are shown for time $t$ = 0, 0.25, 0.5, 0.75, 1.0, 1.13,
1.25, and 1.38, 
respectively.}
\label{fig:11}       
\end{figure*}
 
The shearing sheet is now subjected to two independent impulsive external
potential perturbations of the form given in equation (39) with initial wave
vectors ${\bf k}^{in,1}$ and ${\bf k}^{in,2}$, respectively. As shown in 
Figs.~10 and 11 both perturbations evolve like the single perturbation in the 
previous section.
However, further density waves appear with circumferential wave numbers 
$k_{\rm y}$ as predicted by the sum rule given in equation (19). We
interpret these new density waves, which are delayed relative to the imposed
perturbations, as examples of the onset of the non--linear feedback cycle
predicted theoretically in section (2). In Fig.~10 an example is shown which
started with the initial wave vectors ${\bf k}^{in,1}  = 
(-2, 0.5)\,k_{\rm crit}$ and ${\bf k}^{in,2} = (-3, 1)\,k_{\rm crit}$, 
respectively. At time $t$=0.88 there appears a new weak feature at about
${\bf k} = (0.9, 1.5)\,k_{\rm crit}$. This then couples back with the first
externally triggered wave leading to a strong wave at time $t$ = 1.13 with 
$k_{\rm y} = 1\,k_{\rm crit}$, which travels behind the second externally 
triggered wave. The example in Fig.~11 started with initial wave vectors 
${\bf k}^{in,1} = (-2, 0.5)\,k_{\rm crit}$ and ${\bf k}^{in,2} = 
(-3,1.5)\,k_{\rm crit}$, respectively. At time $t$=0.5 a new weak feature 
appears at about ${\bf k} = (-3, 1.5)\,k_{\rm crit}$. After the second
externally triggered wave was wrapped up very tightly, so that it 
travelled out of the frames shown in Fig.~11, there appears at time $t$ = 1.13
a new strong wave at ${\bf k} = (1, 1)\,k_{\rm crit}$. Non--linear coupling
of density waves with different wave numbers, which led to recurrent density
wave activity, has already been noted by Chiueh \& Tseng (2000) in their 
simulations, although no detailed explanation was offered. They also find the
decline of the amplitudes of density waves at given $k_{\rm y}$ wave number 
which they ascribe to Landau damping. However, since density waves in the
shearing sheet swing around with the shear flow, there is effectively no 
relative streaming of disk particles relative to the waves (cf.~Fig.~3 of 
Toomre 1981), which would lead to Landau damping (Lynden--Bell \& Kalnajs 1972).
Thus in our view hardly any Landau damping is to be expected. The 
density waves die out, because the density contrast is wiped out by the 
epicyclic motions of the stars, when the density wave crests are wrapped up 
tightly by the shear (Julian \& Toomre 1966, Fuchs 2001a).

\section{Summary and conclusion}

The aim of this paper is to study non--linear effects in the dynamical evolution
of an unbounded shearing sheet made of stars, which models a patch of a 
galactic disk. We have analyzed these effects both theoretically and 
by numerical simulations.

First we have investigated theoretically the implication of the non--linear 
term of the Boltzmann equation for the dynamical evolution of the shearing 
sheet. The non--linear term is expressed in
terms of Fourier expansions of the perturbations of the distribution function of
the stars in phase space and the gravitational potential of the sheet,
respectively, and the Boltzmann equation augmented by this term is formally 
solved. It is shown that this further term can be viewed formally as an 
inhomogeneity of the fundamental Volterra integral equation which describes the
dynamical evolution of the shearing 
sheet. The inhomogeneity is given by combinations of the Fourier coefficients
of the perturbations of the distribution function and the gravitational 
potential of the sheet, sampled over the past history  of the evolution of the
sheet. This leads to a feedback cycle for swing amplified density waves. It is
quite different from the effect of non--linearities of the hydrodynamical or
Jeans equations, which have been used alternatively to describe the dynamics 
of the shearing sheet or galactic disks. 

In order to present concrete examples of the non--linear feedback cycle we have
developed a self--consistent field code and have run numerical simulations of 
the dynamical evolution of the
shearing sheet. The SCF method is based on a set of pairwise basis functions
into which the surface density and the gravitational potential of the shearing
sheet are expanded. Each of the pairwise basis functions solve the Poisson
equation. In the case of the shearing sheet these are simply Fourier transforms.
The equations of motion of the particles in the shearing sheet are the 
epicyclic equations of motion, which describe the motion of the particles in the
unperturbed sheet, augmented by the forces due to the fluctuations of the sheet.
In the simulations the shearing sheet turned out to be prone to rapid dynamical
heating. Thus it was necessary to cool the sheet dynamically by adding mass
continously to the sheet during the simulations. 
We have shown that the live shearing sheet responds to impulsive external
potential perturbations by developing shearing density waves. In
particular, by applying simultaneous multiple perturbations we have demonstrated
the appearance of non--linearly induced shearing density waves, exactly as
theoretically predicted.

The feedback cycle found here might well account for the hitherto unexplained 
recurrent shearing spiral instabilities seen in the large scale numerical
simulations of the dynamical evolution of galactic disks.

\acknowledgements{We are indebted to the anonymous referee for comments
that helped to improve the paper. B.F.~thanks E.~Sedlmayr and H.~V\"olk for 
their encouragement to pursue non--linear studies. T.T. acknowledges support 
by a grant from the Alexander--von--Humboldt foundation.}

{}

\appendix
\section{A generalized leap--frog scheme}

We consider a set of ordinary differential equations
\begin{equation}
  \dot{u} = f(u,x), \quad   \dot{x} = g(u,x)\,,
\end{equation}
where $u$ and $x$ may be also vectors. Integrating equations (A.1) by a small
time step $h$ from $t_0$ to $t_0+h$ leads to
\begin{eqnarray}
&&  u(t_0+h) = u_0 + \dot{u}_0 h + \frac{1}{2}\ddot{u}_0 h^2 + O(h^3)
  \nonumber\\ &&
  = u_0 + f(u,x) h + \frac{1}{2} \dot{f}(u,x) h^2 + O(h^3)
  \nonumber\\ &&
  = u_0 + f(u,x) h + \frac{1}{2}h^2 \left[
    \left(f \cdot \frac{\partial }{\partial u}\right)f
    +\left(g \cdot \frac{\partial }{\partial x} \right)f
    \right]  \nonumber \\ && +O(h^3) \,,
  \\ &&
  x(t_0+h) = x_0 + \dot{x}_0h + \frac{1}{2}\ddot{x}_0 h^2 +
  O(h^3)
  \nonumber\\ &&
  = x_0 + g(u,x) h + \frac{1}{2} \dot{g}(u,x) h^2 + O(h^3)
  \nonumber\\ &&
  = x_0 + g(u,x) h + \frac{1}{2}h^2 \left[
    \left( f \cdot \frac{\partial}{\partial u}\right)g +
    \left( g \cdot \frac{\partial}{\partial x}\right)g
    \right] \nonumber \\ && + O(h^3)\,.
\end{eqnarray}
Adopting the principle of the leap--frog algorithm (Hockney \& Eastwood 1988)
we introduce \emph{mid-points},
\begin{equation}
x_{1/2}=x(t_0+h/2), \quad u_{1/2}=u(t_0+h/2) \,, 
\end{equation}
and can write 
\begin{eqnarray}
&&f(u_0,x_{1/2}) = f(u_0,x_0) + \frac{h}{2}\left( g(u_0,x_0) \cdot
    \frac{\partial}{\partial  x}\right) f  + O(h^2)\,,  \nonumber \\ 
&&g(u_{1/2},x_0) = g(u_0,x_0) + \frac{h}{2} \left(f(u_0,x_0) \cdot
    \frac{\partial}{\partial  u}\right) g  + O(h^2)\,.
\end{eqnarray}
It follows then
\begin{eqnarray}
&&  u(t_0+h)
  = u_0 + h f(u_0,x_{1/2}) \nonumber \\ && + 
  \frac{h^2}{2}  \left( f(u_0,x_0) \cdot
    \frac{\partial}{\partial u}\right) f(u_0,x_0) + O(h^3) \nonumber \\ &&
  = u_0 + h f(u_0,x_{1/2}) \nonumber \\ &&
  + \frac{h^2}{2}  \left( f(u_0,x_{1/2}) \cdot
    \frac{\partial}{\partial u}\right) f(u_0,x_{1/2}) + O(h^3)\,,\nonumber \\
&&  x(t_0+h)
  = x_0 + h g(u_{1/2},x_0) \nonumber \\ &&
  + \frac{h^2}{2}  \left( g(u_0,x_0) \cdot
    \frac{\partial}{\partial x}\right) g(u_0,x_0) + O(h^3) \nonumber \\ &&
  = x_0 + h g(u_{1/2},x_0) \nonumber \\ &&
  + \frac{h^2}{2}  \left( g(u_{1/2},x_0) \cdot
    \frac{\partial}{\partial x}\right) g(u_{1/2},x_0) + O(h^3)\,,
\end{eqnarray}
and finally
\begin{eqnarray}
&&  u_{1/2}
  = u_{-1/2} + h f(u_{-1/2},x_0) \nonumber \\ &&
  + \frac{h^2}{2}  \left(
    f(u_{-1/2},x_0) \cdot \frac{\partial}{\partial u}\right)
  f(u_{-1/2},x_0) + O(h^3) \nonumber \\ &&
  x_1
  = x_0 + h g(u_{1/2},x_0) \nonumber \\ &&
  + \frac{h^2}{2}  \left( g(u_{1/2},x_0) \cdot
    \frac{\partial}{\partial x}\right) g(u_{1/2},x_0) + O(h^3)\,.
\end{eqnarray}

The equations of motion (36) of the particles in the shearing sheet can be 
written as a set of first order differential equations,
\begin{eqnarray}
&& \dot{u} = 2 \Omega_0 v + f_{\rm x} \nonumber \\ &&
\dot{v} = 2 B u + f_{\rm y}\nonumber \\ &&
\dot{x} = u \nonumber \\ &&
\dot{y} = v -2 A x \,.
\end{eqnarray}
From equations (A.7) we derive the leap--frog scheme
\begin{eqnarray}
 && u_{1/2} = u_{-1/2} + h\left[2\Omega_0 v_{-1/2}+f_x(x_0,y_0)\right]
 \nonumber \\ &&
  + h^2 \Omega_0 \left[2B u_{-1/2} + f_y(x_0,y_0) \right] +
  O(h^3)\,, \nonumber \\ &&
  v_{1/2} = v_{-1/2} +h \left[ 2B u_{-1/2} + f_y(x_0,y_0) \right] 
  \nonumber \\ &&
  + h^2(A-\Omega_0) \left[2\Omega_0
    v_{-1/2}+f_x(x_0,y_0)\right] +  O(h^3)\,, \nonumber \\ &&
  x_1 = x_0 + h u_{1/2} + O(h^3)\nonumber \\ &&
  y_1 = y_0 + h(v_{1/2} - 2A x_0) - h^2 A u_{1/2} + O(h^3)\,.
\end{eqnarray}
The time step was set to $h$ = 0.03125 in the simulations. Thus the orbits are
integrated with an accuracy of $h^2$ = 0.001. 
\end{document}